\documentclass{cpbtex}
\usepackage{textcomp}
\usepackage{amssymb}
\usepackage[geometry]{ifsym}
\usepackage{MnSymbol}
\usepackage{float}
\usepackage{algorithm}
\usepackage{algpseudocode}
\usepackage{graphicx}
\usepackage{caption}

\usepackage{CJKutf8}

\begin{document}
\begin{CJK*}{UTF8}{gbsn}
\title{1/9 Magnetization Plateau in a Classical Kagome Ising Ferromagnet with Competing Further-Neighbor Interactions}


\author{ Yixin Guan(关益欣)$^{1}$, \, Kan Zhao(赵侃)$^{1}$\thanks{Corresponding author. E-mail:~kan\_zhao@buaa.edu.cn}\ and \ Nvsen Ma(马女森)$^{1}$\thanks{Corresponding author. E-mail:~nvsenma@buaa.edu.cn}\\
$^{1}${School of Physics, Beihang University, Beijing 100191, China}\\ 
}

\date{\today}
\maketitle

\begin{abstract}

The two-dimensional kagome lattice is a paradigmatic platform for exploring geometrically frustrated magnetism. While the nearest-neighbor ferromagnetic Ising model on this lattice is theoretically trivial, competing further-neighbor interactions can reintroduce severe frustration. In this work, we systematically investigate a classical kagome Ising model with ferromagnetic nearest-neighbor ($J_1$) and antiferromagnetic second- ($J_2$) and third-neighbor ($J_3$) couplings using simulated annealing Monte Carlo methods. We demonstrate that while $J_2$ couplings merely suppress the conventional ferromagnetic order, the inclusion of $J_3$ fundamentally reconstructs the low-temperature phase diagram. This extended geometric frustration stabilizes a novel ordered phase characterized by a robust 1/9 magnetization plateau and a massively enlarged $3 \times 3$ magnetic supercell. Crucially, this fractional ordered phase manifests as a stability plateau in the phase diagram, where its critical temperature becomes nearly independent of the coupling strength $J_3$. We also calculate the corresponding static spin structure factor, revealing a distinct $Z_6$-symmetric reciprocal-space signature for experimental identification. Our findings reveal that complex fractional magnetic orders can emerge purely from classical geometric frustration induced by competing extended interactions, providing a distinct mechanism for understanding fractionally ordered states in real frustrated magnets.

\end{abstract}


\section{Introduction}

The study of geometrically frustrated spin systems provides fertile ground for discovering exotic states of matter \ucite {frustration,Sandvik2010,sachdev1991,wangsicheng2026}, with the two-dimensional kagome lattice serving as a paradigmatic architecture \ucite {Chern2011,zhao2020,su2023,zhao2025}. The standard nearest-neighbor Ising model on a kagome lattice with antiferromagnetic interactions is highly frustrated. This geometric frustration prevents the simultaneous satisfaction of all pairwise bonds, resulting in a macroscopic ground-state degeneracy that precludes any finite-temperature phase transition\ucite{frustration,ningcai2025,menghanzhang2023}. In this case, the inclusion of further-neighbor interactions can lift this macroscopic degeneracy, stabilizing a specific ordered phase and inducing transitions in the system\ucite{Chern2012,moessner2009,mizo2017, mila2022,ma2024}. For example, it is well established that adding ferromagnetic next-nearest-neighbor ($J_2$) couplings to a classical disordered ground state induces a two-stage ordering process characterized by consecutive Kosterlitz-Thouless (KT) phase transitions and an intermediate critical phase\ucite{su2023,Chern2012,yujingliu2026}. These various ordered phases selected by longer-range couplings, together with their numerous critical points have been the focus of extensive study\ucite{Chern2012,PhysRevB.93.214410,moessner2009,su2023,plat1018}.

Conversely, if the nearest-neighbor interactions ($J_1$) are purely ferromagnetic, geometric frustration is completely bypassed. The system undergoes a trivial continuous phase transition to a fully aligned ferromagnetic state. However, competing extended exchange interactions, which naturally arise in metallic systems\ucite{dai2022,shiliangli2018} or materials with long-range couplings\ucite{spinice2001,Wei_2020}, may reintroduce frustration into an otherwise unfrustrated background. Yet, the specific scenario in which a trivial ferromagnetic Ising model on the kagome lattice is systematically frustrated by extended antiferromagnetic couplings remains largely unexplored. 

Motivated by the possible rich physics of competing energy scales, we systematically investigate the thermodynamic properties of an extended two-dimensional kagome ferromagnetic Ising model with antiferromagnetic $J_2$ and $J_3$ interactions. In this work, we use Monte Carlo-based simulated annealing to study the classical phase transitions in this extended model. We demonstrate that while next-nearest-neighbor frustration alone merely suppresses the ferromagnetic state, the inclusion of third-neighbor couplings fundamentally reconstructs the low-temperature phase. Specifically, we reveal that the geometric compromise between these extended interactions stabilizes a highly robust ordered phase with a 1/9 fractional magnetization. In the previous study, the magnetic ordered state with 1/9 fractional magnetization on the kagome lattice has been found only in the quantum systems\ucite{ran2023,picot2016,andrade2026,jianxinli2025}. We map the phase boundaries of this fractional state and detail its corresponding $3 \times 3$ crystalline magnetic supercell with $27$ spins, which could be quite different with the fractional plateau phase in the quantum systems\ucite{bodaji2024}. These results highlight how the geometric frustration of longer-range interactions can drive macroscopic fractional symmetry breaking in classical non-bipartite networks. Our findings suggest that complex fractional plateaus do not strictly require exotic quantum mechanisms \ucite{He2024,ran2023}, but can naturally emerge from the geometric competition inherent to metallic systems. We also calculate the structure factor of the classical fractional magnetic order state to provide theoretical guidance for future experiments aimed at identifying such magnetic supercells in crystalline frustrated materials\ucite{experiment1,zhaokan2024np,zhao2026}.

\section{Models and Methods}

\subsection{The 2D Ising Model on the Kagome Lattice}

\begin{figure}[htbp]
\centering
\includegraphics[width=0.6\columnwidth]{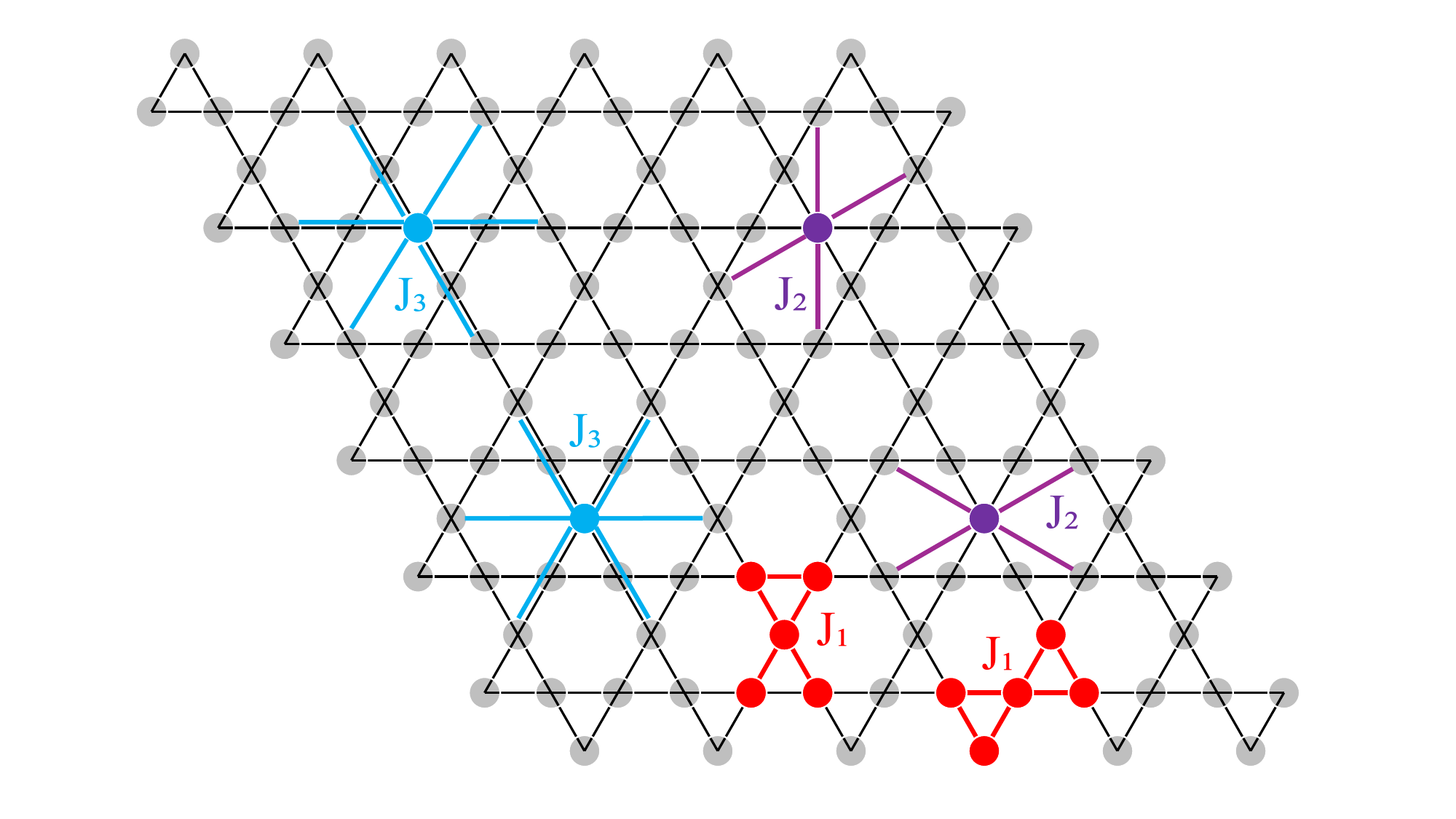}
\caption{Schematic of the two-dimensional kagome lattice illustrating the exchange interactions. The solid red lines, purple lines, and blue lines denote the nearest-neighbor ($J_1$), next-nearest-neighbor ($J_2$), and third-nearest-neighbor ($J_3$) couplings, respectively.}
\label{fig:model}
\end{figure}

In this work, we consider a classical two-dimensional Ising model situated on a kagome lattice with periodic boundary conditions, incorporating exchange interactions up to the third-nearest neighbor. The Hamiltonian of the system is given by:
\begin{equation}
\mathcal{H} = J_1\sum_{\langle ij \rangle} \sigma_i\sigma_j + J_2\sum_{\langle\langle ij \rangle\rangle} \sigma_i\sigma_j + J_3\sum_{\langle\langle\langle ij \rangle\rangle\rangle} \sigma_i\sigma_j
\end{equation}
where $\sigma_i = \pm 1$ represents the Ising spin at site $i$. The first, second, and third sums run over all nearest-neighbor (NN), next-nearest-neighbor (NNN), and third-nearest-neighbor (TNN) spin pairs, respectively. As illustrated in Fig.\ref{fig:model}, the kagome lattice is composed of corner-sharing triangles, with each unit cell containing three sublattices. For a lattice with linear dimension $L$, the total number of spins is $N = 3L^2$. To explore the emergence of frustration-induced states from the trivial ordered states, we set the NN interaction to be ferromagnetic with $J_1 = -1$, which also serves as the energy scale in our simulation. Frustration is introduced through the further-neighbor couplings, which are set to be antiferromagnetic ($J_2 \ge 0$ and $J_3 \ge 0$).
\subsection{Simulated Annealing and Monte Carlo Methodology}

To investigate the thermodynamic properties and phase transitions of this model, we employ the simulated annealing method combined with the standard Metropolis Monte Carlo algorithm\ucite{metropolis}. 

In a single Monte Carlo step (MCS), we attempt to flip every spin in the lattice once. The acceptance of a spin flip is determined by the Metropolis criterion, with the transition probability $p = \min(1, e^{-\Delta E / k_B T})$, where $\Delta E$ is the energy change caused by the proposed flip and $k_B$ is the Boltzmann constant (set to $k_B = 1$ for simplicity).

To ensure statistical reliability and prevent the system from becoming trapped in localized metastable states, which could be a common issue in highly frustrated geometries, we simulate $N_b = 128$ independent bins of the system simultaneously, each initialized with a distinct random seed. For each bin, the system is slowly cooled from a high initial temperature $T_i$ to a final low temperature $T_f$ at a constant cooling rate $\Delta T$. At each temperature step, the system first undergoes $10,000$ MCS for thermal equilibration, followed by an additional $20,000$ MCS for data sampling and statistical averaging.

\subsection{Physical Observables and Finite-Size Scaling}

In this work, we use several macroscopic observables to study the phase transitions of the systems. The first one is the absolute magnetization per spin, which is defined as:
\begin{equation}
\langle |M| \rangle = \frac{1}{N} \left\langle \left| \sum_{i=1}^N \sigma_i \right| \right\rangle, 
\end{equation}
where the angle brackets $\langle \dots \rangle$ denote the thermal average over the sampling MCS.  We also calculate the specific heat $C$ , which can be got from the fluctuations of the internal energy $E$ as 
\begin{equation}
C = \frac{1}{N k_B T^2} (\langle E^2 \rangle - \langle E \rangle^2). 
\end{equation}

To determine the phase transition temperatures $T_c$ and critical exponents, we calculate the dimensionless fourth-order Binder cumulant $U_2$:
\begin{equation}
U_2 = 1 - \frac{\langle M^4 \rangle}{3\langle M^2 \rangle^2}. 
\label{eq:binder}
\end{equation}
According to finite-size scaling (FSS) theory, in the vicinity of a continuous phase transition, the Binder cumulant follows the universal scaling form:
\begin{equation}
U_2(T, L) = \tilde{U}( t L^{1/\nu})
\end{equation}

where $\tilde{U}$ is a universal scaling function and $\nu$ is the critical exponent associated with the correlation length and $t=\frac{T - T_c}{T_c} $. At the critical temperature $T = T_c$, the argument of the scaling function becomes zero, rendering $U_2$ independent of the system size $L$. Consequently, the $U_2(T)$ curves plotted for various lattice sizes should cross at one point as the critical temperature $T_{c}$. 

But in practice, the crossing points for different sizes differ from each other, resulting from the corrections to the finite size scaling, where
\begin{equation}
U_2(T, L) = \tilde{U}( tL^{1/\nu})(1+aL^{\omega})
\end{equation}
with $\omega$ the leading correction exponent. Thus, the crossing points extracted from finite-size systems scales toward the thermodynamic critical point $T_c(L \to \infty)$ according to the power law:
\begin{equation}
T_c(L) = T_c(\infty) + a L^{-\omega}
\label{eq:scaling}
\end{equation}
where $a$ is a non-universal constant. 

\begin{figure}[htbp]
\centering
\includegraphics[width=1.0\columnwidth]{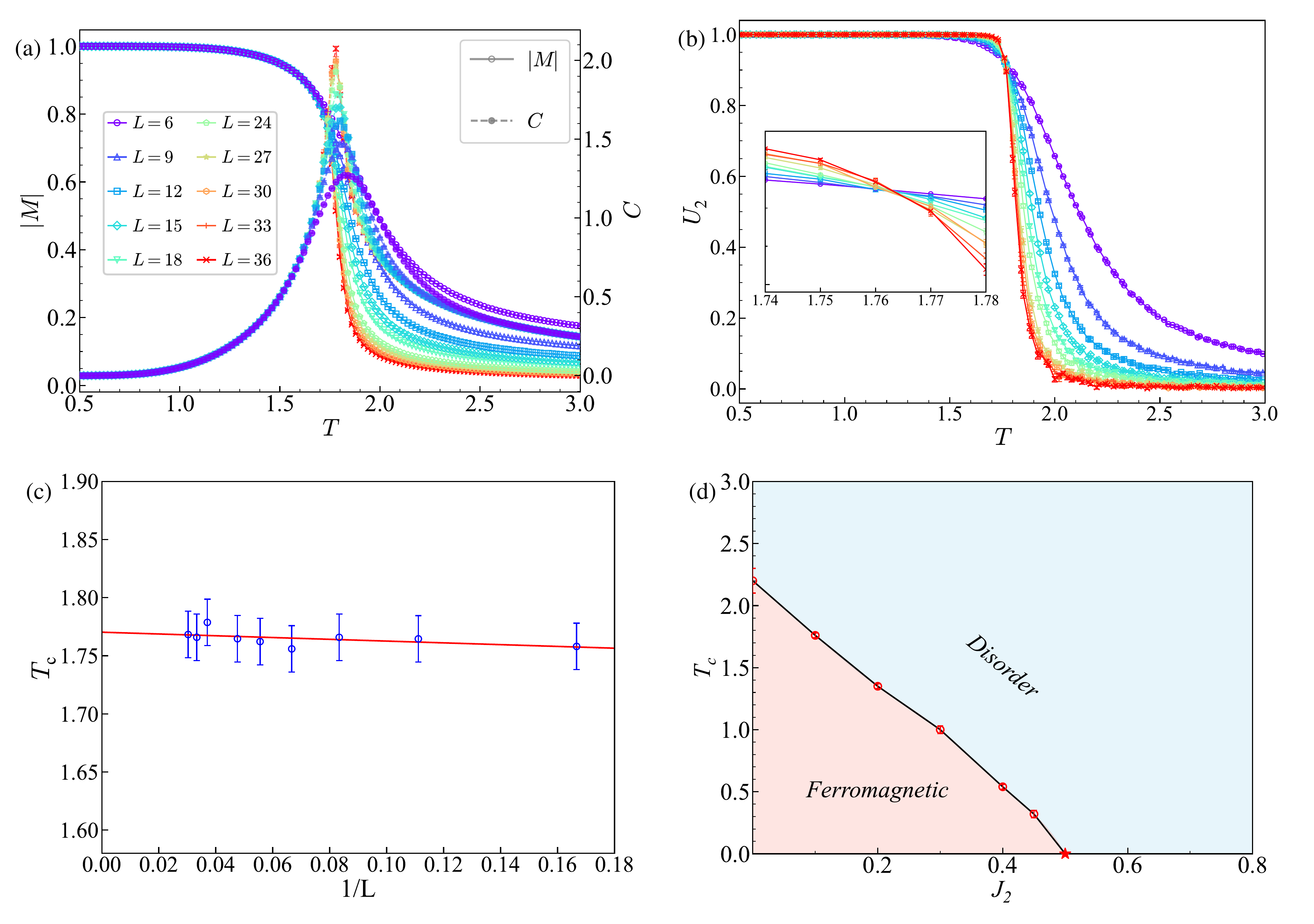}
\caption{Thermodynamic phase transition and classical phase diagram for the kagome Ising model with $J_1 = -1$ and $J_2 > 0$. (a) Temperature dependence of absolute magnetization $|M|$ and specific heat $C$ for a representative coupling of $J_2 = 0.1$ across varying lattice sizes $L$. (b) The Binder cumulant $U_2$ as a function of temperature for different $L$. The inset provides a magnified view of the critical region with crossing points for various sizes, shifting minimally. (c) The thermodynamic critical temperatures $T_c(L)$ extracted from the intersection points of (b) are plotted against $1/L$. The $y$-intercept of the fitting curve with the scaling form in Eq.~\ref{eq:scaling} implies that $T_c(\infty) \approx 1.768$. (d) The $T_c - J_2$ phase diagram. The red circles represent the thermodynamic critical temperatures $T_c(\infty)$ obtained by repeating the aforementioned finite-size scaling procedure across various values of $J_2$. For $J_{2}=0.5$, the magnetic order cannot be found when $T\rightarrow 0$ with no phase transitions at finite temperature. The red star implies that the boundary extrapolates to $T=0$ at $J_2=0.5$ as an approximate value due to the discrete resolution of our $J_2$ parameter sampling.}

\label{fig:j1j2_baseline}
\end{figure}

\section{Results}

\subsection{Ferromagnetic Kagome Ising Model with Next-Nearest-Neighbor Interactions}

Before investigating the effects of the third-nearest-neighbor coupling, we first establish the baseline thermodynamic behavior of the kagome Ising model governed solely by nearest-neighbor (NN) and next-nearest-neighbor (NNN) interactions. We set a purely ferromagnetic NN coupling ($J_1 = -1$) and introduce an antiferromagnetic NNN coupling ($J_2 > 0$). In the absence of $J_2$, the model trivially undergoes a continuous phase transition belonging to the $Z_2$ universality class into a conventional ferromagnetic ordered state at low temperatures.

To determine how the introduction of $J_2$ perturbs this ordered state, we track the temperature dependence of the absolute magnetization $|M|$, the specific heat $C$, and the Binder cumulant $U_2$. Figure 2(a) illustrates the macroscopic observables as a function of temperature for a representative antiferromagnetic coupling of $J_2 = 0.1$ across various lattice sizes $L$. As the system cools, the magnetization sharply increases from zero, accompanied by a divergent peak in the specific heat, signaling the onset of ferromagnetic order.  The temperature dependence of  $U_2$ for different sizes in Fig. 2(b) cross at the same point, presenting little influence of the scaling corrections. We also extract these crossing points and perform a finite-size scaling extrapolation plotted against $1/L$, as presented in Fig. 2(c),  yielding a critical temperature of $T_c(L \to \infty) \approx 1.768$ for $J_2 = 0.1$. All crossing points of $T_{c}(L)$ stay at the same value considering the error, which is similar to the ferromagnetic Ising model in the two-dimensional square lattice.

By systematically varying $J_2$ and repeating this finite-size scaling procedure, we map the classical $T_c - J_2$ phase diagram for this regime, presented in Fig. 2(d). The introduction of the antiferromagnetic $J_2$ interaction acts to destabilize the ferromagnetic alignment dictated by $J_1$. Consequently, the critical temperature $T_c$ decreases almost linearly as $J_2$ increases. This suppression continues until $J_2 \approx 0.5$, beyond which the ferromagnetic phase is entirely destroyed by the strong NNN interactions, giving way to a highly degenerate disordered regime down to $T=0$.


\subsection{Emergence of the 1/9 Ordered Phase Driven by Third-Neighbor Couplings}


To investigate how the inclusion of third-nearest-neighbor (TNN) interactions alters the macroscopic magnetic order, we fix the baseline parameters to $J_1 = -1$ and $J_2 = 0.2$ with a ferromagnetic ground state and introduce a frustrating antiferromagnetic TNN coupling $J_3 > 0$. The first clear signature of a novel ordered state comes into being when analyzing the temperature dependence of the physical observables.

\begin{figure}[htbp]
\centering
\includegraphics[width=1.0\columnwidth]{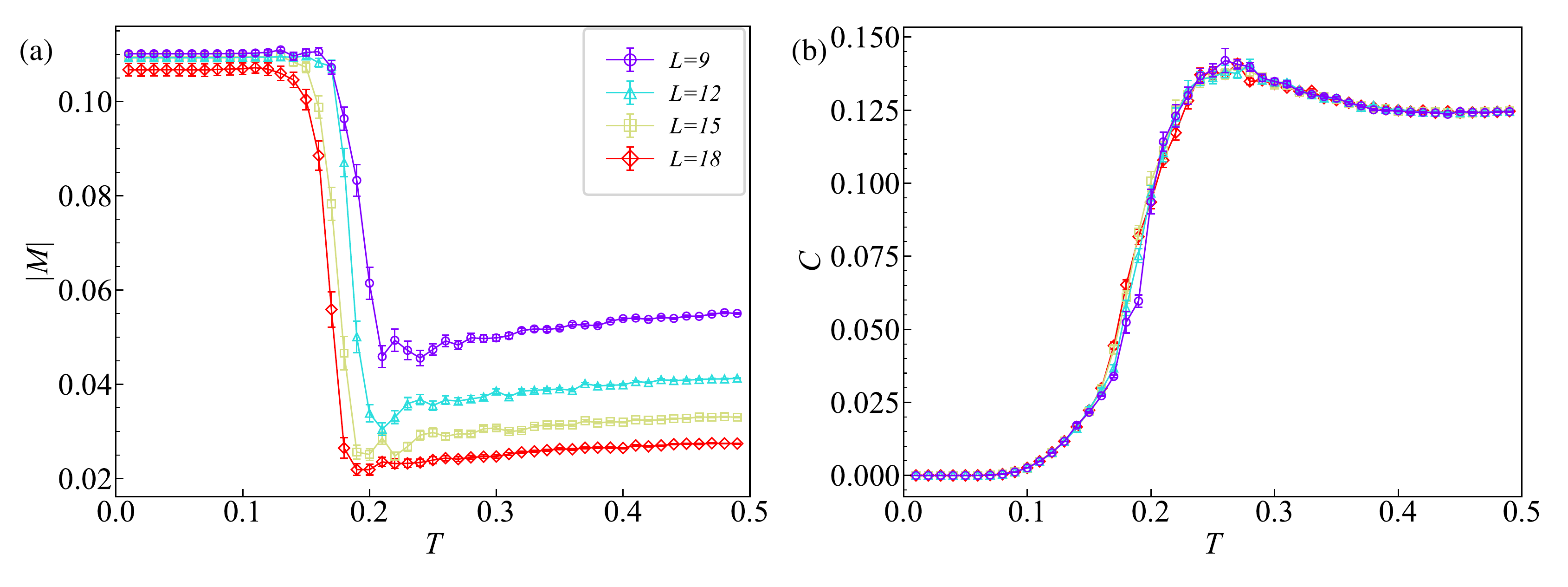}
\caption{Thermodynamic signatures of the fractional phase transition driven by third-nearest-neighbor frustration. The baseline parameters are fixed to $J_1 = -1$ and $J_2 = 0.2$, with $J_3 = 0.22$. (a) Temperature dependence of the absolute magnetization $|M|$ across varying lattice sizes $L$. As $L$ increases, the low-temperature macroscopic magnetization systematically converges toward a robust fractional value of $|M| = 1/9$. (b) Temperature dependence of the specific heat $C$, displaying a single, pronounced peak at $T \approx 0.25$ that perfectly coincides with the onset of the 1/9 ordered phase.}
\label{fig:j3_observables}
\end{figure}

Figure \ref{fig:j3_observables} illustrates the magnetization and the specific heat as a function of temperature for a representative TNN coupling of $J_3 = 0.22$ across various lattice sizes $L$. As the system cools, the specific heat in Fig.~\ref{fig:j3_observables} exhibits a single, pronounced peak at $T \approx 0.25$, indicating a direct thermodynamic phase transition. However, unlike the previously discussed $J_1-J_2$ case, the system does not enter a conventional ferromagnetic state at low temperatures. Instead, as shown in Fig. \ref{fig:j3_observables}(a), magnetization drops rapidly concurrently with the specific heat peak, stabilizing at a finite but much lower value for temperatures $T \lesssim 0.25$. Crucially, a distinct finite-size scaling trend is observed: as the lattice size $L$ increases, the low-temperature magnetization curves systematically converge toward a specific fractional value of $|M| = 1/9 \approx 0.111$. This convergence toward a fractional value for the macroscopic order parameter strongly indicates the emergence of a magnetic ordered ground state driven by the extended geometric frustration introduced by $J_{3}$.

\begin{figure}[htbp]
\centering
\includegraphics[width=1.0\columnwidth]{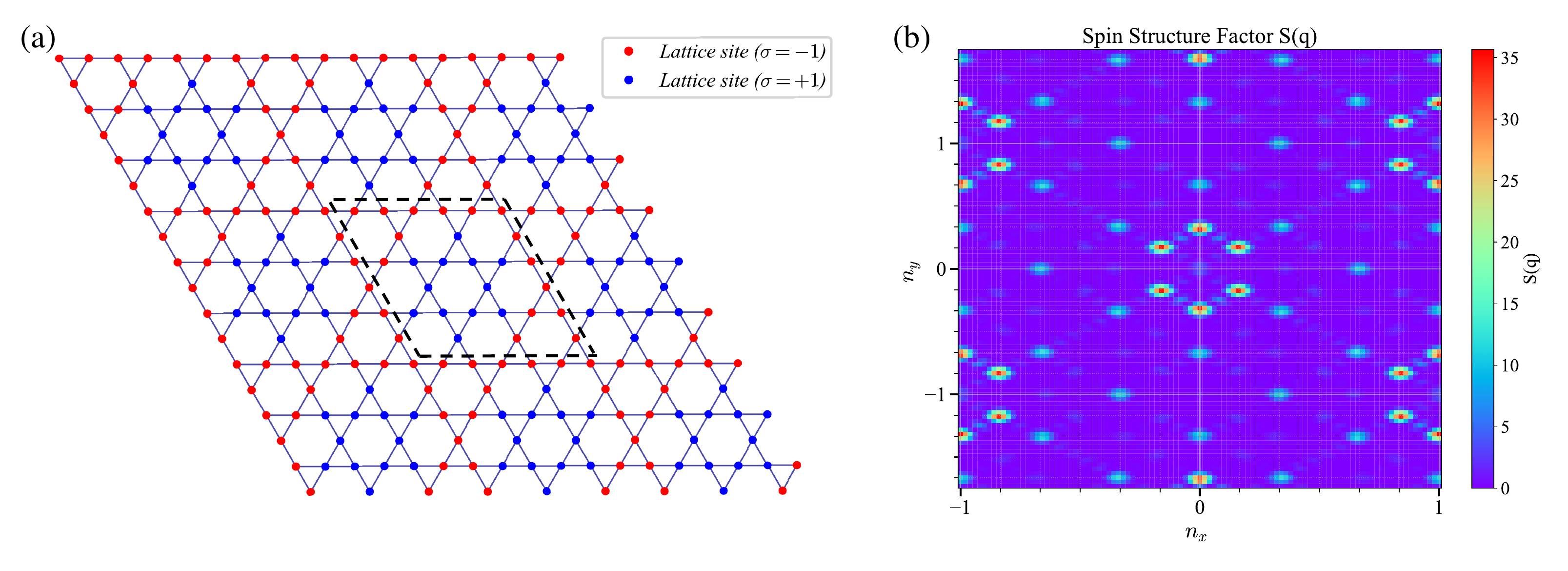}
\caption{Microscopic magnetic structure and symmetry of the 1/9 magnetization plateau state. (a) Real-space spin configuration illustrating the spontaneous translational symmetry breaking into a $3 \times 3$ enlarged magnetic supercell, as indicated by the dashed black lines. Red and blue circles represent spins with $\sigma = -1$ and $\sigma = +1$, respectively. (b) The corresponding static spin structure factor $S(\mathbf{q})$ in the reciprocal space. The emergence of sharp superlattice Bragg peaks at fractional reciprocal lattice vectors (forming a six-fold symmetric pattern) rigorously confirms the long-range $3 \times 3$ crystalline magnetic order. The structure factor is shown as a function of the wave vector $\mathbf{n} = (n_x, n_y)$, where the components $n_x$ and $n_y$ are expressed in units of the combinations of reciprocal lattice vectors $\mathbf{b}_1 + \mathbf{b}_2$ and $\mathbf{b}_1 - \mathbf{b}_2$.}
\label{fig:1_9_structure}
\end{figure}

So we continue to examine the microscopic magnetic structure of the ordered state with a fractional magnetization. The real-space spin configuration presented in Fig.~\ref{fig:1_9_structure} deep within the ordered phase implies that the system stabilizes into a massive magnetic unit cell, forming a $3 \times 3$ supercell containing nine original kagome unit cells (27 spins in total). In this configuration, most spins align ferromagnetically to satisfy the strong $J_1$ bonds. However, to relieve the geometric frustration imposed by the extended couplings, a specific, periodically arranged minority of spins flip to satisfy the antiferromagnetic $J_2$ and $J_3$ bonds. This precise spatial compromise perfectly yields the $\langle |M| \rangle = 1/9$ value.

To rigorously confirm this long-range ordering and provide a direct, testable prediction for future elastic neutron scattering experiments, we calculate the static spin structure factor $S(\mathbf{q}) = \frac{1}{N} \langle |\sum_i \sigma_i e^{i \mathbf{q} \cdot \mathbf{r}_i}|^2 \rangle$. As shown in Fig. 4(b), the structure factor is plotted as a function of the wave vector $\mathbf{n} = (n_x, n_y)$, where the components $n_x$ and $n_y$ are expressed in units of the combinations of the reciprocal lattice vectors $\mathbf{b}_1 + \mathbf{b}_2$ and $\mathbf{b}_1 - \mathbf{b}_2$.

The structure factor in the ordered state presents a hexagram pattern with a distinct six-fold symmetric ($Z_6$). It is instructive to compare this reciprocal-space signature to other known ordered phases on the kagome lattice, such as the ordered spin ice state. While the ordered spin ice also exhibits a $Z_6$ symmetry, its dominant Bragg peaks are located at entirely different characteristic wave vectors. Therefore, the specific peaks in Fig. \ref{fig:1_9_structure}(b) not only confirms the enlarged $3 \times 3$ magnetic unit cell , but also serves as a distinct, identifying experimental fingerprint for this novel  1/9 ordered phase in reciprocal space.

\subsection{Phase Boundaries and the Stability Plateau of the 1/9 Ordered Phase}

Having established the thermodynamic and microscopic properties of the 1/9 ordered phase, we now map its boundaries within the broader parameter space. Because this fractional state emerges as a compromise between competing energy scales, it is crucial to understand the macroscopic interplay between the next-nearest-neighbor ($J_2$) and third-nearest-neighbor ($J_3$) interactions.

\begin{figure}[htbp]
\centering
\includegraphics[width=1.0\columnwidth]{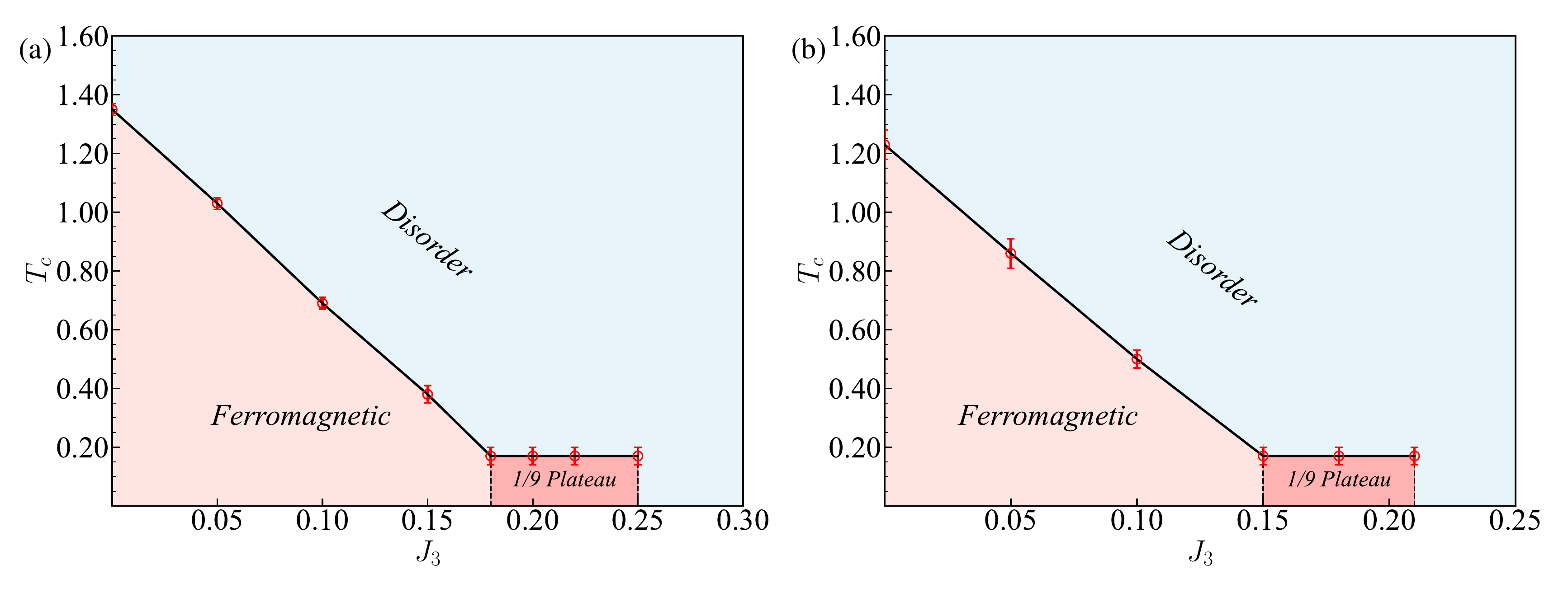}
\caption{Classical phase diagrams illustrating the stability plateau of the 1/9 ordered phase. (a) $T_c$ versus $J_3$ for a fixed coupling of $J_2 = 0.20$. The $T_c$ boundary abruptly flattens into a plateau at $J_3 \gtrsim 0.18$, marking the robust onset of the 1/9 ordered phase. (b) $T_c$ versus $J_3$ for $J_2 = 0.25$. The increased $J_2$ frustration shifts the emergence of the plateau to a lower coupling of $J_3 \gtrsim 0.15$.}
\label{fig:j3_phasediagrams}
\end{figure}

To visualize this competition, we map the $T_c - J_3$ phase diagrams for two fixed values of $J_2$. As shown in Figure \ref{fig:j3_phasediagrams}(a) for $J_2 = 0.20$, increasing $J_3$ initially suppresses the critical temperature of the ferromagnetic phase in a nearly linear fashion. However, once $J_3$ reaches a critical threshold ($J_3 \gtrsim 0.18$), the phase boundary abruptly flattens. In this region, the system enters the 1/9 ordered phase, and the critical temperature becomes almost entirely insensitive to further increases in $J_3$, maintaining a nearly constant $T_c \approx 0.2$. This flat phase boundary forms a distinct, robust macroscopic stability plateau in the phase diagram.

Figure \ref{fig:j3_phasediagrams}(b) illustrates the phase diagram when the next-nearest-neighbor frustration is increased to $J_2 = 0.25$. The increased $J_2$ further destabilizes the nearest-neighbor ferromagnetic alignment, suppressing the ferromagnetic boundary a little bit more rapidly. More importantly, we also find a stability plateau once the system enters the 1/9 ordered phase, even with a lower coupling value of $J_3 \gtrsim 0.15$ in this case. It shows that as the antiferromagnetic $J_2$ increases, the required strength of $J_3$ to stabilize the 1/9 ordered phase decreases. We believe that the 1/9 ordered phase presents as a robust stability plateau across a continuous range of both $J_2$ and $J_3$, meaning that it could be a highly stable thermodynamic phase rather than a sensitive mathematical anomaly. Because real metallic systems often possess competing extended interactions whose magnitudes naturally decay with distance, this purely classical geometric stabilization mechanism suggests that such fractionally ordered phases and their corresponding plateaus could be observable in experimental frustrated magnetic systems.

\section{Conclusion}

In summary, we have systematically investigated the thermodynamic properties and phase transitions of the classical two-dimensional kagome Ising model with competing extended exchange interactions. With simulated annealing Monte Carlo methods, we demonstrated that while next-nearest-neighbor antiferromagnetic couplings ($J_2$) merely suppress the conventional ferromagnetic order induced by nearest-neighbor bonds ($J_1$), the introduction of third-nearest-neighbor antiferromagnetic couplings ($J_3$) fundamentally alters the thermodynamic landscape. 

The central finding of this work is the discovery of a highly symmetric fractional ground state as the 1/9 magnetization ordered phase. Through careful analysis of specific heat and low-temperature spin configurations, we showed that the system undergoes a thermodynamic phase transition into this state. Microscopically, the phase is stabilized by the spontaneous breaking of translational symmetry, forming a massive $3 \times 3$ magnetic supercell with $27$ spins that perfectly compromises between the competing $J_1$, $J_2$, and $J_3$ energy scales while preserving $Z_6$ symmetry.

Furthermore, by mapping the precise phase boundaries, we find that as the next-nearest-neighbor interactions increase, the critical strength of $J_3$ required to stabilize the 1/9 ordered phase decreases. In the phase diagram, this phase manifests as a flat stability plateau, where the critical temperature becomes insensitive to further increases in $J_3$. The existence of this robust stability plateau across a continuous parameter space indicates that the 1/9 ordered phase is a stable thermodynamic reality rather than a mathematical anomaly. 

Ultimately, these results highlight that complex fractional magnetic orders—often attributed exclusively to strong quantum fluctuations or entanglement in Heisenberg models—can emerge purely from classical geometric frustration. Because real metallic frustrated magnets frequently exhibit competing extended exchange interactions whose magnitudes naturally decay with distance, the purely geometric mechanism identified in this classical Ising model provides a vital theoretical foundation for understanding and searching for novel fractionally ordered phases in real-world two-dimensional materials.

\section{Acknowledgements}

We thank Hua Chen for helpful discussion on the
possible ordered phases on the kagome lattice. The work was supported by the National Key R$\&$D Program of China (Grants No. 2023YFA1406003), the National Natural Science Foundation of China (Grant No.12474139 and No.12274015), the Fundamental Research Funds for Central Universities and the Beijing Natural Science Foundation
(Grant No. JQ24012).
\bibliographystyle{cpb}
\bibliography{kagome_ref}

\end{CJK*}
\end{document}